\documentclass[pra,twocolumn,showpacs,letterpaper,showpacs,superscriptaddress]{revtex4}
\usepackage{amsfonts}
\usepackage{amsmath}
\usepackage{amssymb}
\usepackage{graphicx}
\newcommand{\be}{\begin{equation}}
\newcommand{\ee}{\end{equation}}
\newcommand{\bea}{\begin{eqnarray}}
\newcommand{\eea}{\end{eqnarray}}

\preprint{ }

\begin{document}

\title{Inertial forces and dissipation on accelerated boundaries}
\author{C.D. Fosco}
\affiliation{Centro At\'omico Bariloche and Instituto Balseiro, Comisi\'on Nacional de Energ\'\i a At\'omica}

\author{F.C. Lombardo}
\affiliation{Departamento de F\'\i sica {\it Juan Jos\'e
Giambiagi}, FCEyN UBA, Facultad de Ciencias Exactas y Naturales,
Ciudad Universitaria, Pabell\' on I, 1428 Buenos Aires, Argentina}

\author{F. D. Mazzitelli}
\affiliation{Departamento de F\'\i sica {\it Juan Jos\'e
Giambiagi}, FCEyN UBA, Facultad de Ciencias Exactas y Naturales,
Ciudad Universitaria, Pabell\' on I, 1428 Buenos Aires, Argentina}

\date{today}

\begin{abstract}
We study dissipative effects due to inertial forces acting on matter fields
confined to accelerated boundaries in $1+1$, $2+1$, and $3+1$ dimensions.
These matter fields describe the internal degrees of freedom of `mirrors'
and impose, on the surfaces where they are defined, boundary conditions on
a fluctuating `vacuum' field.  We construct different models, involving
either scalar or Dirac matter fields coupled to a vacuum scalar field, and
use effective action techniques to calculate the
strength of dissipation. In the case of massless Dirac fields, the results
could be used to describe the inertial forces on an accelerated graphene
sheet. 
\end{abstract}


\maketitle

\section{Introduction}

In what is one of the most startling manifestations of the quantum nature
of the electromagnetic (EM) field, the Dynamical Casimir Effect (DCE)
predicts that, in the presence of an accelerated mirror, the quantum vacuum
evolves to an excited state (i.e., one with a non-vanishing number of
photons). As a consequence, the mirror will experience a dissipative force.
Both the dissipation and the number of created photons are rather small for
a single accelerated mirror. However, the effect can be enhanced in
electromagnetic cavities with time dependent length or, more generally,
with time dependent electromagnetic properties (see the
reviews~\cite{reviews}, and references therein).

In this letter we consider an effect that, although of a different nature
than the DCE, may nevertheless arise under fairly similar conditions: acceleration 
of the mirrors. The main difference with the DCE is that we shall consider the `inertial'
dissipation that results from the transfer of energy to the matter fields inside 
the mirror, rather than to the EM field.

Effects due to the inertial forces in accelerated media are known, and usually 
described in terms of induced
time-dependent polarizations or currents, like in the Barnett and
Stewart-Tolman effects~\cite{bar1}.  More generally, there will be 
inertial forces as long as there is energy transfer from the collective 
coordinates of the mirrors, that specify their position and shape as a function of time, to their internal degrees of freedom. 

Contrary to the vacuum friction effects considered up to now
\cite{reviews}-\cite{particle}, the inertial effects described in this
letter will exist even when the internal degrees of freedom are not coupled
to the EM field. 
Our aim in this work is to evaluate these effects using
specific models for the matter fields, allowing for mirrors with a quite
 general space and time dependence. 

\section{The model}

Let us begin by considering a scalar model, described by an Euclidean
action $S$ with the structure:
\begin{equation}\label{eq:defs}
S \;=\; S_f \,+\, S_m \,+\, S_I\;\;,
\end{equation}
where $S_f$ describes the vacuum field dynamics in the absence of mirrors.
It will be assumed to be a free field theory in  $d+1$ dimensions ($d=1$,
$2$ or $3$).  $S_m$ is the free part of the matter field action, i.e., it
ignores the coupling to the vacuum field; this coupling is instead
contained in the last term, $S_I$. 

Both $S_m$ and $S_I$ have support on ${\mathcal M}$, the $d$-dimensional
spacetime manifold swept by $\Sigma(\tau)$, the space occupied by the
mirrors at the time $\tau$, in the course of their time evolution. 

The vacuum field $\varphi$ is assumed to be a scalar, equipped with
the action 
\begin{equation}\label{eq:defsvac}
S_f(\varphi) \;=\; \frac{1}{2} \int d^{d+1}x \; \partial_\mu \varphi(x)
\partial_\mu \varphi(x) \;.
\end{equation} 
In this scalar model, the matter degrees of freedom are described by $\xi$,
also a real scalar field, but living on ${\mathcal M}$, a curved
$d$-dimensional manifold. To make this more explicit, we introduce $d$
variables: $\sigma^0$, $\sigma^1$, \ldots $\sigma^{d-1}$, such that
$\sigma^0$ is the time coordinate; ${\mathcal M}$ can then be parametrized
by $d+1$ functions $y^\mu$, $\mu = 0,\,1,\,\ldots,d$, as follows:
\begin{equation}
{\mathcal M}: \;\;\; \sigma \; \longrightarrow \; y^\mu(\sigma^0,\ldots, \sigma^{d-1})
\end{equation} 
such that $y^0 = \sigma^0 \equiv x^0 \equiv \tau$ and $y^j = y^j(\tau ,
\sigma^1, \ldots, \sigma^{d-1})$ ($j=1,\ldots d$), is a parametrization of
$\Sigma(\tau)$ (for each $\tau$).  Note that ${\mathcal M}$ may denote
either one or more than one mirror.  When dealing with more than one
mirror, we shall introduce an alternative, more explicit notation, obtained
by using a different parametrization for each connected piece in ${\mathcal M}$. 

We assume the scalar matter field action to be given by:
\begin{eqnarray}\label{eq:defsm} S_m(\xi; {\mathcal M}) &=& \frac{1}{2}
\int d^d\sigma \sqrt{g(\sigma)} \big[ g^{\alpha\beta}(\sigma)
\partial_\alpha \xi(\sigma)  \partial_\beta
\xi(\sigma)\nonumber \\
&+& \mu^2 \xi^2(\sigma) \big] \;,
\end{eqnarray} 
where $g_{\alpha\beta}(\sigma)$ is the induced metric on ${\mathcal M}$:
\begin{equation}
g_{\alpha\beta}(\sigma) = \partial_\alpha y^{\mu}(\sigma) \partial_\beta
y^{\mu}(\sigma) \;\;,\;\;\;\;\; \alpha=0,1,\ldots, d-1 \;,
\end{equation}
$g(\sigma)$ its determinant, and $\mu$ the mass of the field.

For the $S_I$ piece, on the other hand, we shall consider a term of the form:
\begin{equation}\label{eq:defsi1}
S_I(\varphi, \xi; {\mathcal M}) \;=\; - i \,
\zeta \, \int d^d \sigma \,  \sqrt{g(\sigma)}
\, \xi(\sigma) \, \varphi\big[ y(\sigma) \big] \;,
\end{equation}
where $\zeta$ denotes the coupling constant.

The action $S$ can be considered as a toy model for the interaction between
the EM field ($\varphi$ in the toy model) and charged degrees of freedom
living on the mirrors (described by $\xi$), although, as it shall be
evident below, it is unnecessary to have a coupling between the internal
degrees of freedom and the vacuum field for dissipation to arise.  It is
nevertheless interesting to consider models where that coupling exists, in
order to compare `inertial' and DCE dissipations. 

The scalar nature of the matter field is appropriate to the field
configuration describing an `internal' (charge) degree of freedom.  We
shall also consider the case of a Dirac-like matter field, but we shall not
dwell with vector-like matter fields, which could account for phonon-like
excitations, and certainly also exhibit inertial dissipation. 

To take into account the effects of the internal degrees of freedom, we introduce
$\Gamma_m(\varphi)$, the effective action due to the matter fields, defined as follows:
\begin{equation}\label{eq:defgammam}
e^{-\Gamma_m(\varphi;{\mathcal M})} \;\equiv\; 
\int {\mathcal D} \xi \; e^{-S_m(\xi; {\mathcal M}) - S_I(\varphi, \xi; {\mathcal M})}  \;,
\end{equation}
which also encompasses the dependence on the geometry of ${\mathcal M}$.

Because of their quite different properties, we split $\Gamma_m$ into two 
contributions, as follows:
\begin{equation}
\Gamma_m(\varphi;{\mathcal M}) \;=\; 
\Gamma_i({\mathcal M}) 
\,+\, 
\Gamma_b(\varphi; {\mathcal M}) 
\end{equation}   
where 
$$
\Gamma_i({\mathcal M}) \equiv \Gamma_m(\varphi;{\mathcal M})|_{\varphi=0},$$ 
$$\Gamma_b(\varphi; {\mathcal M})=
\Gamma_m(\varphi;{\mathcal M}) \,-\, \Gamma_m(\varphi;{\mathcal
M})|_{\varphi=0}.$$

Each one of these terms is relevant to different effects:
$\Gamma_i({\mathcal M})$ can induce dissipative effects, while
$\Gamma_b(\varphi;{\mathcal M})$ is responsible for the emergence of
boundary conditions on $\varphi$. Our focus shall be the calculation of 
$\Gamma_i({\mathcal M})$, although we shall also derive the form of 
$\Gamma_b(\varphi;{\mathcal M})$ for the sake of completeness. 

It is quite straightforward to see that:
\begin{equation}\label{eq:gms}
\Gamma_i({\mathcal M}) \;=\; \frac{1}{2} \, {\rm Tr} \ln {\mathcal K}  
\end{equation}
where ${\mathcal K}$ is the operator
\begin{equation}
{\mathcal K} \;=\;  - \partial_\alpha \Big[g^{1/2}
g^{\alpha\beta}\partial_\beta \Big] + g^{1/2} \mu^2 \;=\; g^{1/2} 
\, \big( -\Delta_{\mathcal M} + \mu^2 \big) \;,
\end{equation}
and $\Delta_{\mathcal M} \equiv g^{-1/2} \partial_\alpha \Big[g^{1/2}
g^{\alpha\beta}\partial_\beta \Big] $ is the Laplacian corresponding to the
induced metric on ${\mathcal M}$. Note that the operator ${\mathcal K}$ is
Hermitian for the scalar product $(f_1,f_2) \,=\, \int d^d\sigma \,
\big(f_1(\sigma)\big)^* f_2(\sigma)$. 

Besides, the boundary interaction term $\Gamma_b(\varphi;{\mathcal M})$, is 
given by: 
\begin{eqnarray}\label{eq:gbs}
\Gamma_b(\varphi;{\mathcal M}) &=& \frac{\zeta^2}{2} \, 
\int d^d \sigma g^{1/2}(\sigma) \int d^d \sigma' g^{1/2}(\sigma')
\nonumber\\
&\times&  \varphi\big[y(\sigma)\big] {\mathcal K}^{-1}(\sigma,\sigma')
\varphi\big[y(\sigma')\big] \;.  
\end{eqnarray}

Expressions (\ref{eq:gms}) and (\ref{eq:gbs}), properly interpreted,
are valid for one or more than one mirrors. If there is just one mirror,
their interpretation is immediate: one needs to know a parametrization of
its surface, then all the objects involved in (\ref{eq:gms}) and
(\ref{eq:gbs}) are derived from it and the free scalar field propagator. 
When there are $N > 1$ mirrors, we introduce a discrete index $a$, and use
a different notation for the functions that parametrize each mirror:
$y^\mu_a$, $a=1,\ldots, N$ (which can be, for example, restrictions of the
parametrization of ${\mathcal M}$) . Then, since the action for the matter
fields is localized on each mirror, we see that the effective action 
becomes the sum of the contributions due to each mirror: 
\begin{equation}
\Gamma_i({\mathcal M}) = \sum_{a=1}^N \Gamma_i({\mathcal M}_a)   
\;,\;\;\;
\Gamma_b(\varphi;{\mathcal M}) = \sum_{a=1}^N \Gamma_b(\varphi;{\mathcal
M}_a) \;,  
\end{equation}
where ${\mathcal M}_a$ is the world-volume of each mirror.

In other words, quantum effects of the matter fields are, for this model,
additive with respect to the mirrors. Thus, in the rest of this paper we
shall only calculate $\Gamma_i$ and $\Gamma_b$ for single mirrors.  Before
presenting results of the evaluation of those contributions for different
numbers of dimensions, we check that, in the limit when the microscopic
degrees of freedom do not propagate, the boundary interaction term reduces
to the standard one. Indeed, keeping just the leading term in a large-$\mu$
expansion:
\begin{equation}
{\mathcal K}^{-1}(\sigma,\sigma') \; \sim \; \mu^{-2} \; 
g^{-1/2}(\sigma) \, \delta(\sigma-\sigma') 
\end{equation}
so that for the leading behaviour of the boundary interaction term we have
the following expression:
\begin{equation}
\Gamma_b(\varphi;{\mathcal M}) \;\sim\;  \frac{\lambda}{2} \, 
\int d^d \sigma g^{1/2}(\sigma) \varphi^2\big[y(\sigma)\big] 
\end{equation}
where $\lambda \equiv \big(\frac{\zeta}{\mu}\big)^2$.  This is the usual
$\delta$-like potential that has been used to analyze the Casimir effect with 
imperfect boundary conditions \cite{deltap}.

\section{No inertial dissipation in $1+1$ dimensions}

To gain some insight into the previous expressions, let us first evaluate 
the effective action for a case where the outcome will  be that 
there will be no inertial dissipation, namely,
$d=1$. Here the vacuum field is a massless real scalar field (in $1+1$
dimensions) while $\xi$ is a single quantum mechanical degree of freedom.
For the model we consider, $\xi$ cannot experience inertial forces,
since it is not spatial (its values belong to an internal space), 
and moreover it is only a function of time. 

The spacetime manifold ${\mathcal M}$ is one-dimensional, a plane curve
which we parametrize as follows: $\tau \to (\tau, q(\tau))$, where $\tau$
is the time and $q(\tau)$ the position of the mirror.  
The $S_m$ and $S_I$ terms may therefore be written in the form:
\begin{eqnarray}\label{eq:defsm1}
S_m(\xi; {\mathcal M}) &=& \frac{1}{2} \int d\tau \Big[ e^{-1}(\tau)
\big( \frac{d\xi(\tau)}{d\tau} \big)^2  + \mu^2 e(\tau) \xi^2(\tau) \Big]
\nonumber\\
S_I(\varphi,\xi; {\mathcal M}) &=& - i \zeta \int d\tau  \, e(\tau) \,
\xi(\tau) \, \varphi(\tau,q(\tau)) 
\end{eqnarray} 
where $e(\tau) \equiv \sqrt{g(\tau)}$, and 
\mbox{$g(\tau) \equiv 1 + \dot{q}^2(\tau)$}. 
Then we note that, by performing the reparametrization: $\tau \to s$, such
that $ds = e(\tau) d\tau$, we get:
\begin{eqnarray}\label{eq:defsm2}
S_m(\xi; {\mathcal M}) &=& 
\frac{1}{2} \int ds \Big[ \big( \frac{d\widetilde{\xi}(s)}{ds} \big)^2  
+ \mu^2  \widetilde{\xi}^2(s) \Big]
\nonumber\\
S_I(\varphi,\xi; {\mathcal M}) &=& -i \zeta \, \int ds  \,
\widetilde{\xi}(s) \, \widetilde{\varphi}(s,\tilde{q}(s)) 
\end{eqnarray} 
where we have introduced the notations:
\begin{equation}
\widetilde{\xi}(s) \equiv \xi[\tau(s)]\;,\;\;
\widetilde{\varphi}[s, \widetilde{q}(s)] \equiv \varphi[\tau(s),
\widetilde{q}(s)] \;,
\end{equation}

Integrating out the matter field (we assume its path integral to be
invariant under reparametrizations~\footnote{The reparametrization used, on
the other hand, is always well defined since $e \neq 0$. We assume, having
in mind the real time version, that the speed of the mirror is always
smaller than $c$.}), we see that in the effective action term $\Gamma_i$, the dependence
on $e(\tau)$ (hence on $q(\tau)$) is completely erased. Thus no dissipative
effects from this origin may arise in $d=1$, as advanced. 

Regarding the $\Gamma_b$ term, we find that:
\begin{equation}  
\Gamma_b(\varphi;{\mathcal M}) \,=\, \frac{\zeta^2}{2} \, 
\int ds \int ds' \, {\widetilde\varphi}(s,{\widetilde q}(s))
\, \Delta(s-s') \,      
{\widetilde\varphi}(s',{\widetilde q}(s'))      
\end{equation}
with $\Delta(s-s') \,=\, e^{- \mu |s-s'|}/2 \mu$. This boundary interaction
term can be also written in a Fourier representation, so that it produces a
frequency dependent coupling between the vacuum field and the mirror: 
\begin{equation}  
\Gamma_b(\varphi;{\mathcal M}) \,=\, \frac{\zeta^2}{2} \, 
\int \frac{d\omega}{2\pi} \, \frac{1}{\omega^2 + \mu^2} \, 
\big|{\mathcal F}({\widetilde\varphi})(\omega)\big|^2    \;, 
\end{equation}
where ${\mathcal F}$ denotes Fourier transform with respect to the $s$
variable. We note that the outcome of this study amounts to the property
that the $\Gamma_b$ term is, essentially, a reparametrization invariant
interaction, which is tantamount to using a relativistic invariant term. For
the non-dynamical limit, this is precisely one of the cases we have studied in
\cite{Fosco:2007nz}. Indeed, when the matter degrees of freedom do not
propagate,
\begin{eqnarray}  
\Gamma_b(\varphi;{\mathcal M}) &\sim & \frac{\lambda}{2} \, 
\int ds \big[ {\widetilde\varphi}(s,{\widetilde q}(s)) \big]^2 \nonumber \\
&=& \frac{\lambda}{2} \, \int d\tau \sqrt{1 + \dot{q}^2(\tau)}
\big[\varphi(\tau,q(\tau))\big]^2 \;,
\end{eqnarray}
with $\lambda \equiv \frac{\zeta^2}{\mu}$, which is identical to equation
(33) of~\cite{Fosco:2007nz}.

In the general, nonlocal case, we can still find the effective action
$\Gamma[q(\tau)]$, for a single mirror, obtained by performing now the
functional integration over the vacuum field:
\begin{equation}
e^{-\Gamma[q(\tau)]} \,=\, \int {\mathcal D}\varphi \, e^{ - S_f(\varphi) -
\Gamma_b(\varphi;{\mathcal M}) }\;.
\end{equation}
 To second order in the mirror departure from its average position, we get:
\begin{equation}
\Gamma[q(\tau)]= - \int_{-\infty}^{+\infty} d\tau e(\tau)
\int_{-\infty}^{+\infty} d\tau' e(\tau') q(\tau) F_s(\tau-\tau') q(\tau') 
\end{equation}
where:
\begin{equation}
F_s(\tau-\tau') = \int \frac{d\omega}{2\pi} e^{i \omega (\tau-\tau')}
 {\widetilde F}_s(\omega) \;,\;{\widetilde F}_s(\omega) = 
{\widetilde F}(\omega) - {\widetilde F}(0),
\end{equation}
and:
\begin{equation}
{\widetilde F}(\omega) \,=\, \frac{1}{4} \, \int \frac{d\nu}{2\pi} \, 
\big[ \frac{(\nu + \omega)^2 + \mu^2}{\zeta} + \frac{1}{2 |\nu + \omega|} \big]^{-1} |\nu| \;. 
\end{equation}
Although the integral cannot be evaluated exactly, it is clear that its 
UV behaviour is much improved with respect to the one corresponding to a local $\Gamma_b$, like in~\cite{Fosco:2007nz}. The physical reason  for that is that the nonlocal $\Gamma_b$ introduces a kind of cutoff frequency (of order $\mu$) above which the mirror becomes ineffective to impose the boundary conditions. 

\section{$2+1$ dimensions: Inertial dissipation on a moving string}

 The situation is more interesting regarding dissipation when $d=2$, since
the space dependence of the matter field allow for the action of inertial
forces. Besides, if $\mu=0$, it is possible to derive a closed-form
expression for $\Gamma_m({\mathcal M})$. Indeed, this object is formally
identical to the effective action for a massless real scalar field ($\xi$)
in two dimensions, which can be found exactly, and it is nontrivial only if
$R$, the scalar curvature of ${\mathcal M}$ is different from $0$.  Thus,
before writing the explicit form of the result, we find conditions for $R$
to be different from $0$.  

Using ${\mathbf y}$ to denote the two spatial components of the parametric
form of ${\mathcal M}$, namely, ${\mathbf y} = {\mathbf
y}(\sigma^0,\sigma^1)$ ($\sigma^0
\equiv \tau$, $\sigma^1\equiv\sigma$), we
find, after a quite straightforward calculation, that:
\begin{equation}
R \;=\; \frac{2}{g^2} \; A
\end{equation}
where $g$ is the determinant of the induced metric, and:
\begin{eqnarray}
A &=& 
(\partial_1y^1)^2 \, \partial_0^2 y^2 \, \partial_1^2 y^2 \,-\, 
\partial_1y^1 \, \partial_1^2 y^1 \, \partial_0^2 y^2 \, \partial_1y^2  \nonumber\\
&-& \partial_0^2 y^1 \, \partial_1 y^1 \, \partial_1 y^2 \, \partial_1^2 y^2 
\,+\,  (\partial_1y^2)^2 \, \partial_0^2 y^1 \, \partial_1^2 y^1 \nonumber\\
&-& (\partial_1 y_1)^2 \, (\partial_0\partial_1 y^2)^2 -   (\partial_1
y^2)^2 \, (\partial_0\partial_1 y^1)^2 \nonumber\\
&+& 2 \partial_1 y^1 \, \partial_1 y^2 \, \partial_0\partial_1 y^1 \,\partial_0\partial_1 y^2 \;,
\end{eqnarray}
where $y^1$ and $y^2$ are the two spatial components of ${\mathbf y}$. A
simpler expression for $R$ may be written when the surface can be written
as a Monge patch (which is then necessarily open), that is, when ${\mathcal
M}$ can be parametrized as:
\begin{equation}
y^0 = \tau \;,\;\;\;
y^1 = \sigma\;,\;\;\;
y^2 = y(\tau,\sigma)\, .
\end{equation}
Note that $y(\tau,\sigma)$ describes a moving open string in $2+1$ dimensions.
In this case we have
\begin{equation}\label{eq:rmong}
R \;=\; \frac{2}{g^2} \; F\,\, , 
\end{equation}
with
\begin{equation}\label{eq:deff}
F \,=\, \left| \begin{array}{cc}
\partial_\tau^2 y & \partial_\tau \partial_\sigma y \\
\partial_\sigma\partial_\tau y & \partial_\sigma^2 y 
\end{array} 
\right|
\;\;,\;\;
g = 1 + (\partial_\tau y)^2 + (\partial_\sigma y)^2 \;.
\end{equation}
A necessary and sufficient condition for $F$ to vanish can be obtained:
\begin{equation}
F = 0 \;\; \leftrightarrow \;\; \exists \alpha \in {\mathbb R} \;\; / \; 
( \partial_\tau + \alpha \partial_\sigma) y(\tau,\sigma) = 0 \;.  
\end{equation}
This includes  (since $\alpha$ can also be $0$ or $\infty$), the cases of an $F$ 
which depends only on $\tau$ or only on $\sigma$. We may write then all
solutions to the condition above, for $R$ to be zero (and $\Gamma_m$ trivial) as 
follows:
\begin{equation}
y = f(\tau - \alpha^{-1} \sigma) \;,
\end{equation}
(for any $\alpha$). Thus, there will be no effect if the
evolution of the mirror has the form of a packet that evolves undisturbed,
with any speed (eventually zero). That is to say, they are solutions of the wave
equation with wave speed $\alpha$ (arbitrary), and with a definite
chirality: namely, they should either move to the left or to the right, but
cannot be a combination of both. 
The absence of dissipative effects for this kind of evolutions is
certainly related to the Lorentz invariance of the model.

On the other hand, combining both left and right movers, 
one certainly can have
an $F \neq 0$. Thus, what is perhaps the simplest  example of a configuration where $F
\neq 0$ corresponds to standing waves. 

We now recall the form of the (exact) effective action for a massless 
real scalar field in $1+1$ dimensions \cite{Polyakov}, to write:
\begin{equation}
\Gamma_m ({\mathcal M}) \;=\; - \frac{1}{96\pi} \, \int d^2x \, R
\frac{1}{\Delta} R \;.
\end{equation}
We shall, in what follows, only consider the massless matter field case.
The reason is that for a massive field the corresponding contribution is
strongly suppressed, unless the curvature is big in comparison with a scale
set by the mass $\mu$. In such a case, however, the massless field case
should be a good approximation. Moreover, in the massive case there are
also local terms in the curvature, but their locality makes them irrelevant
to dissipative effects.

Of course, inserting the expression corresponding to a Monge Patch, one
gets a more explicit form. Doing that, one may look for the $\Gamma_m$ that
results from the lowest order contribution, in a derivative expansion of 
$y(\tau,\sigma)$: 
\begin{eqnarray}
\Gamma_m & \simeq & \frac{1}{24\pi} \int d\tau d\sigma 
 \int d\tau' d\sigma' 
F(\tau, \sigma)\Delta_0^{-1}(\tau-\tau',\sigma-\sigma') \nonumber \\ &\times & F(\tau',
\sigma'). 
\end{eqnarray} 
where $\Delta_0^{-1}$ is the free  propagator (which has the usual
logarithmic form), and $F$ has been defined in (\ref{eq:deff}).

The pole in the free propagator is responsible for the existence of a
nontrivial dissipation. Indeed, we may Fourier transform, and rotate to 
real time, to obtain the imaginary part of $\Gamma_m$:
\begin{eqnarray}
{\rm Im}\, \big[\Gamma_m] &\simeq & \frac{1}{96\pi} \, \int \frac{dk_1}{2\pi}
\frac{1}{|k_1|} \; \Big[ \big|\widetilde{F}(k_0=|k_1|,k_1)\big|^2
\nonumber \\ &+&  \big|\widetilde{F}(k_0=-|k_1|,k_1)|^2 \Big] \;,
\label{Im2d}
\end{eqnarray} 
where $\widetilde{F}(k_0,k_1)$ is the Fourier transform of $F$. Note that
the imaginary part sees the arguments of this function on-shell. On the
other hand, $F$ is quadratic in $y$ and therefore the dissipation is quartic
in the deformation of the string.

\section{$3+1$ dimensions: Inertial dissipation on a moving mirror}

We conclude our study of the scalar field model with the  $d=3$ case. Here,
it is not possible to find an exact effective action for an arbitrary
manifold ${\mathcal M}$ (even  when $\mu=0$).
The effective action can, however, be computed approximately, in an
expansion in powers 
of the curvature. One has, up to the second order~\cite{Avramidi:1990je}:
\begin{eqnarray}
\Gamma_i({\mathcal M}) &\simeq& - \frac{1}{64 \times 2^{3/2}} \; \int d^3\sigma
\sqrt{g(\sigma)} \Big[ a_1 R_{\alpha\beta} (-\Delta)^{-\frac{1}{2}}
R_{\alpha\beta}\nonumber\\
&+& a_2 R(-\Delta)^{-\frac{1}{2}}  R \Big] +  \Gamma_{local}\;,
\label{scalareff}
\end{eqnarray}
where $a_1=-1$ and $a_2=1/8$. $\Gamma_{local}$  is a divergent contribution constructed 
with the induced metric and its derivatives, and we will assume that these divergences
are absorbed into appropriate counterterms in the classical action.
Since local terms are not relevant for the dissipative effects considered here,
we will omit $\Gamma_{local}$ in what follows.

The nonlocal kernel $(-\Delta)^{-\frac{1}{2}}$ can be formally defined using the 
integral representation \cite{dln07}
\begin{equation}
(-\Delta)^{-\frac{1}{2}}=\frac{2}{\pi} \int_0^{+\infty}
dm \;  \frac {1}{-\Delta +m^2} \; ,
\label{integralrep}
\end{equation}
in terms of the two-point function of a massive scalar field $(-\Delta +m^2)^{-1}$. 
 
Consistently with the expansion of the effective action in powers of the curvature,  we shall  
analyze the nonlocal part of  $\Gamma_i({\mathcal M})$ under the assumption 
that the mirror is almost flat, and  that the time dependence of its surface is smooth.
This means small departures from a flat surface almost flat
manifold ${\mathcal M}$. To implement this in a quite
straightforward way, we first assume that the parametrization used is of
the form:
\begin{equation}
y^\mu = \left\{ \begin{array}{cl}
\sigma^\mu&\;\;{\rm for} \; \mu = 0,\,1,\, 2 \\
y(\sigma^0,\sigma^1,\sigma^2) &\;\;{\rm for} \, \mu=3 \;. 
\end{array}
\right.
\end{equation} 
Then it follows that $g_{\alpha\beta}= \delta_{\alpha\beta} +
h_{\alpha\beta}$, with $h_{\alpha\beta} \equiv 
\partial_\alpha y \partial_\beta y$.
We then perform an expansion in the deformation of the manifold, i.e.,
around the $\delta_{\alpha\beta}$ metric: $g_{\alpha\beta}
=\delta_{\alpha\beta} + h_{\alpha\beta}$, assuming that $ h_{\alpha\beta}$
is small.  To find the lowest non-trivial contribution in an expansion in
$h$, we note that: 
\begin{eqnarray}
\Gamma_i({\mathcal M}) &\simeq&  \frac{1}{64 \times 2^{3/2}} 
\int d^3\sigma R_{\alpha\beta} (-\Delta_0)^{-\frac{1}{2}} \nonumber \\
&\times & [\delta_{\alpha\rho}\delta_{\beta\sigma}
-\frac{1}{8} \delta_{\alpha\beta}\delta_{\rho\sigma}
]
R_{\rho\sigma}\; ,
\end{eqnarray}
where  $R_{\alpha\beta}$ is the Ricci tensor expanded to its lowest non-trivial 
order in $h_{\alpha\beta}$ and $ (-\Delta_0)^{-\frac{1}{2}}$ is the nonlocal kernel
for a flat spacetime. 

To linear order in $h_{\alpha\beta}$ we have
\begin{equation}
R_{\alpha\beta}\simeq\frac{1}{2}(\partial_\alpha\partial_\beta h_{\gamma\gamma}+
\partial_\gamma\partial_\gamma h_{\alpha\beta} -\partial_\alpha\partial_\gamma h_{\gamma\beta}-
\partial_\beta\partial_\gamma h_{\gamma\alpha})
\label{Riccilin}
\end{equation}
Using this equation, it is straightforward to write the Ricci tensor in terms of the function $y(\sigma)$. For instance we have
\begin{eqnarray}
R_{00}&\simeq & \partial_0\partial_1y\,\partial_0\partial_1y
+\partial_0\partial_2y\,\partial_0\partial_2y
-\partial_0\partial_0y\,\partial_1\partial_1y \nonumber \\
&-& \partial_0\partial_0y\,\partial_2\partial_2y\; ,
\label{Riccilin00}\end{eqnarray}
and similar expressions for the other components. 

As in the two-dimensional case, 
one can readily show that all 
components of the Ricci tensor vanish when the evolution of the mirror
is described by a packet that moves with constant velocity in any direction, i.e.
\mbox{$y=f(\sigma^0-\alpha^{-1}\sigma^i)$}. Therefore,  
the manifold ${\mathcal M}$ is flat for this kind of evolutions, since in $d=3$ the Riemann
tensor is entirely determined by the Ricci tensor. Once more, simple evolutions 
that produce dissipative effects are standing waves. 

The dissipative effects induced by  the matter degrees of freedom are related to the imaginary part
of the effective action in Minkowski spacetime. Rotating to real time, and introducing
Fourier transforms for the flat propagator and the Ricci tensor we obtain
\begin{eqnarray}
{\rm Im}\, \big[\Gamma_i\big] &\simeq & {\rm Im} \left[ \frac{1}{32\pi 2^{3/2}} \int_0^\infty dm
 \int \frac{d^3k}{(2\pi)^3}\right. \nonumber \\ &\times & \left. \frac{\widetilde{R}_{\alpha\beta}(k)
 \widetilde{R}_{\rho\sigma}(-k) [\eta^{\alpha\rho}\eta^{\beta\sigma}
-\frac{1}{8} \eta^{\alpha\beta}\eta^{\rho\sigma}
] }{(-k_0^2+k_1^2+k_2^2+m^2-i\epsilon)}\right].
\end{eqnarray} 
Introducing the notation
\begin{equation}
\vert \widetilde{ B}(k)\vert^2= \widetilde{R}_{\alpha\beta}(k)
 \widetilde{R}_{\rho\sigma}(-k) [\eta^{\alpha\rho}\eta^{\beta\sigma}
-\frac{1}{8} \eta^{\alpha\beta}\eta^{\rho\sigma}
] \; ,
\end{equation}
and performing the integration over $k_0$, we find
\begin{eqnarray}
{\rm Im}\, \big[\Gamma_i\big] &\simeq  &\frac{1}{2^{3/2}64} \int_0^\infty dm\,
 \int \frac{d^2k}{(2\pi)^2}\nonumber \\ &\times &\frac{\vert \widetilde{ B}(k_0=\omega_{\mathbf k},{\mathbf k})\vert^2+\vert \widetilde{ B}(k_0=-\omega_{\mathbf k},{\mathbf k})\vert^2}{\omega_{\mathbf k} }
\label{Imeff}\end{eqnarray} 
with $\omega_{\mathbf k}=\sqrt{k_1^2+k_2^2+m^2}$. The structure of this
result is similar to that of the two-dimensional case (\ref{Im2d}), except
for the additional integral in the parameter $m$, used to have a 
suitable integral representation of the nonlocal kernel $(-\Delta_0^2)^{-\frac{1}{2}}$.  

It is worth mentioning that a case in which the matter field is a massless
Dirac field $\psi$ (in $2+1$ dimensions) leads to very similar results.
Assuming the matrices $\gamma^a$ ($a=1, 2, 3$) for this field to be in a
reducible representation of Clifford's algebra, chosen in such a way that
the parity anomaly is cancelled, the matter field action $S_m$ is given by:
\begin{equation}
S_m({\bar\psi},\psi;{\mathcal M}) \;=\; \int d^3\sigma \sqrt{g(\sigma)} \,
{\bar\psi}(\sigma) \, D \,  \psi(\sigma)
\end{equation} 
where $D = \gamma^a e^\mu_a(\sigma) D_\mu$, $e^\mu_a$ is the dreibein
and $D_\mu = \partial_\mu - \frac{i}{4} \omega^a_{b \mu} \sigma_{ab}$. 
Here, $\omega^a_{b \mu}$ denotes the spin connection (determined by the dreibein
and Christoffel symbol) and $\sigma_{ab} \equiv \frac{i}{2} [\gamma_a, \gamma_b]$. 
The result for $\Gamma_i$ may also be read from~\cite{Avramidi:1990je}, and it is given by
(\ref{scalareff}) with the only difference  being in the numerical values
of the coefficients $a_1$ and $a_2$, that we will not need in
what follows. Thus, all the estimates below, as well as
the conditions for no dissipation hold true for this model as well. Note
that the Dirac field model might be relevant to models describing graphene
surfaces \cite{graphene}.

We can obtain an estimation of ${\rm Im}\, \big[\Gamma_i\big]$ for the
scalar and Dirac fields, by assuming that the dynamics of the mirror is
described by a standing wave of the form 
\begin{equation}
 y(\sigma^0,\sigma^1,\sigma^2) = y_0 \cos(\Omega\sigma^0) \cos\left(\sigma^1/L\right),
\label{stand}\end{equation}
where we have denoted by $y_0$ the amplitude of the oscillations, $\Omega$
its frequency and $L$ the typical distance between nodes in the direction
of $\sigma^1$. For simplicity, consider evolutions of the mirror that are
invariant along $\sigma^2$. From (\ref{Riccilin}) and (\ref{Riccilin00}),
we can calculate the Ricci tensor. For example, we can estimate the Fourier
transform of (\ref{Riccilin00}), which can be written as
\begin{eqnarray}\label{eq:r00} &{\tilde R}_{00}&(k) = 
-\frac{4\pi^3 y_0^2\Omega^2}{L^2} \delta(k_2)\big[ \delta(k_1)( \delta(k_0 - 2
\Omega) \nonumber \\ &+& \delta(k_0 + 2 \Omega) )  + \delta(k_0)
(\delta(k_1 - 2/L) + \delta(k_1 + 2/L))\big].\nonumber \end{eqnarray} Similar
expressions can be found for the 11-component of the Ricci tensor, and it
is easy to 
see that $R_{10}=R_{01}=0$. 
 
In order to have a rough estimation of the magnitude of ${\rm Im}\, \big[\Gamma_i\big]$ when 
the mirror's dynamics is approximated by Eq.(\ref{stand}), we need to perform integrations 
in Eq.(\ref{Imeff}). After integrating in $k_1$ and $k_2$, we get
\begin{equation}
 {\rm Im}\, \big[\Gamma_i\big] \sim\frac{y_0^4\Omega^3T \Sigma}{L^4} ,
\end{equation}
where $T$ is the total time during which the mirror is moving, and $\Sigma$ 
denotes the total surface of the (unperturbed) mirror. The presence of these coefficients is 
a byproduct of having considered 
a delocalized standing wave as the one in (\ref{stand}). It is worth noting that,
for this particular motion of the mirror, there is no threshold for the dissipation
effects.  

Restoring $\hbar$ and the propagation velocity ($v_F$) of the matter fields 
inside the layer,
the estimated magnitude of ${\rm Im}\, \big[\Gamma_i\big]$ becomes
\begin{equation}
\frac{ {\rm Im}\, \big[\Gamma_i\big]}{T\Sigma} \sim \frac{\hbar y_0^4\Omega^3}{v_F^2L^4}
\,\, .
\end{equation}

We may compare this result with the one obtained for one perfect mirror
 oscillating according with Eq.~(\ref{stand}). In this case,
 the DCE produces an imaginary part in the effective action
 as long as $\Omega>c/L$ \cite{kardargolestanian98}: 
\begin{equation}
\frac{ {\rm Im}\, \big[\Gamma^{\rm DCE}\big]}{T\Sigma} \sim \frac{\hbar y_0^2\Omega^5
(1-\frac{c^2}{L^2\Omega^2})^{5/2}}{c^4}
.\end{equation}
When $\Omega$ is above threshold, the ratio 
between both effects is approximately given by
\begin{equation}
 \frac{{\rm Im}\, \big[\Gamma_i\big]}{{\rm Im}\, \big[\Gamma^{\rm DCE}\big]} \sim \left(\frac{y_0}{L}\right)^2
\left(\frac{c}{v_F}\right)^2 
\left(\frac{c}{\Omega L}\right)^2
\left(1-\frac{c^2}{L^2\Omega^2}\right)^{-5/2}
\, .
\label{compa}
\end{equation}
This ratio may be bigger or smaller than one depending on the value of the different
parameters. It is interesting to remark that while 
the inertial dissipation is quartic in the amplitude
$y_0$, the dissipation associated to the DCE is quadratic. 
Moreover, for the particular motion (\ref{stand}), the inertial effects
do not have a frequency threshold, while the DCE does.

In Eq.~(\ref{compa}) we compared the inertial dissipation with the DCE for a perfect mirror.
Note, however, that, being independent of the coupling to the vacuum field, 
the inertial effect may be more relevant than the DCE in cases where that coupling
is rather weak, namely, when the mirrors are far from perfect.

It is also worth noting that the result mentioned above, that a deformation
traveling undistorted (with a constant direction) does not produce
inertial dissipation, may be relevant to the DCE. Indeed,
in the case of having more than one mirror with such deformation, 
 the configuration is free of inertial dissipation,
 while the DCE does not vanish. 

Finally, we wish to point out that we have considered here the effect due
to inertial forces just for one matter field; in a real medium many different contributions can
appear, coming from the excitation of different internal degrees of freedom.
All of them will contribute to the imaginary part, with different
parameters and kinematical factors.

\section*{Acknowledgements}
C.D.F. thanks CONICET, ANPCyT and UNCuyo for financial support. The work of
F.D.M. and F.C.L was supported by UBA, CONICET and ANPCyT.

\end{document}